\documentclass[useAMS,usenatbib]{mn2e}
\bibpunct{(}{)}{;}{a}{}{,}
\usepackage{graphicx}
\usepackage{latexsym}
\usepackage{amssymb}
\usepackage{multicol}

\title{Collisions and close encounters involving massive main sequence stars}

\author[James. E. Dale, Melvyn. B. Davies]{James. E. Dale$^{1}$\thanks{E-mail: Jim.Dale@astro.le.ac.uk (JED)},
Melvyn. B. Davies$^{2}$\\
$^{1}$Department of Physics and Astronomy, University of Leicester, University Road, Leicester, LE1 7RH, UK\\
$^{2}$Lund Observatory, Box 43, SE-221 00 Lund, Sweden}

\begin{document}

\pagerange{\pageref{firstpage}--\pageref{lastpage}} \pubyear{2002}

\maketitle

\label{firstpage}

\def\mnras{MNRAS}
\def\aaps{A\&ASS}
\def\aj{AJ}
\def\apj{ApJ}
\def\aap{A\&A}
\def\apjl{ApJL}
\def\apjs{ApJS}
\def\aplett{ApL}
\def\aas{A\&AS}
\def\nat{Nature}
\def\pasj{PASJ}
\def\one{$1$ M$_{\odot}$ star}
\def\nine{$9$ M$_{\odot}$ star}
\def\ones{$1$ M$_{\odot}$ stars}
\def\nines{$9$ M$_{\odot}$ stars}
\def\oneg{$1$ M$_{\odot}$ star }
\def\nineg{$9$ M$_{\odot}$ star }
\def\onesg{$1$ M$_{\odot}$ stars }
\def\ninesg{$9$ M$_{\odot}$ stars }

\begin{abstract}
We study close encounters involving massive main sequence stars and the evolution of the exotic products of these encounters as common--envelope systems or possible hypernova progenitors. We show that parabolic encounters between low-- and high--mass stars and between two high--mass stars with small periastrons result in mergers on timescales of a few tens of stellar freefall times (a few tens of hours). We show that such mergers of unevolved low--mass stars with evolved high--mass stars result in little mass loss ($\sim0.01$ M$_{\odot}$) and can deliver sufficient fresh hydrogen to the core of the collision product to allow the collision product to burn for several million years. We find that grazing encounters enter a common--envelope phase which may expel the envelope of the merger product. The deposition of energy in the envelopes of our merger products causes them to swell by factors of $\sim100$. If these remnants exist in very densely-populated environments ($n\gtrsim10^{7}$ pc$^{-3}$), they will suffer further collisions which may drive off their envelopes, leaving behind hard binaries. We show that the products of collisions have cores rotating sufficiently rapidly to make them candidate hypernova/gamma--ray burst progenitors and that $\sim0.1\%$ of massive stars may suffer collisions, sufficient for such events to contribute significantly to the observed rates of hypernovae and gamma--ray bursts.

\end{abstract}

 
\section{Introduction}
Stellar collisions are an important physical process. Although stars are very small in comparison to the typical sizes of stellar clusters, long stellar lifetimes, high number densities, gravitational focusing, or a combination of these factors can permit a significant fraction of the populations of some stellar systems to undergo collisions.\\
\indent Globular clusters, with number densities of $\sim10^{4}$ pc$^{-3}$, velocity dispersions of $\sim10$ km s$^{-1}$ and long---lived low--mass stellar populations, have long been known to host stellar collisions. \cite{1976ApL....17...87H} estimate that $40\%$ of the stars in some globular clusters have suffered collisions. The remnants are thought to appear on cluster colour--magnitude diagrams as part of the cluster's population of blue stragglers, although it does not appear that all blue stragglers are products of collisions \citep{2004MNRAS.349..129D}. The collisional formation of these objects has been studied hydrodynamically by \cite{1987ApJ...323..614B}, \cite{1993ApJ...412..593L}, \cite{2002ApJ...568..939L}, \cite{1997ApJ...477..335S} and \cite{2002MNRAS.332...49S}.\\
\indent Work by \cite{1998MNRAS.298...93B} showed that star clusters initially experience a phase of contraction which can drive the stellar density in the core to extreme values ($\sim10^{8}$ pc$^{-3}$). Under these conditions, encounter timescales can be sufficiently short to allow short--lived high--mass stars to experience collisions. Older stellar systems with high (but not extreme) number densities, such as Orion,  will also rarely host collisions involving massive stars. The appearance and properties of merger remnants depend crucially upon the degree of mixing induced by the collision. \cite{1987ApJ...323..614B}, \cite{2002ApJ...568..939L}, \cite{1997ApJ...477..335S} and \cite{2002MNRAS.332...49S} all studied encounters between low--mass main sequence stars and, with the exception of  \cite{1987ApJ...323..614B}, found that mixing was minimal. It is important to determine if this is also the case in encounters involving high--mass stars, since the merger may deliver fresh hydrogen to the core of the massive star, potentially rejuvenating it.\\
\indent \cite{2002ApJ...576..899P} proposed that the extreme densities thought to exist in some very young stellar systems may trigger a runaway merger process, leading to the formation of an intermediate--mass black hole. These objects have been posited as an explanation for Ultra--Luminous X-ray sources (e.g. \cite{2000PASJ...52..533T}) and as the building blocks for the supermassive black holes residing in the cores of many galaxies. \cite{freitag_gurkan} conducted Monte Carlo simulations of star clusters experiencing core collapse and hence achieving sufficient stellar densities for runaway collisions to occur. To include the effects of stellar collisions in stellar cluster simulations, \cite{2005MNRAS.358.1133F} performed a very large suite of smoothed-particle hydrodynamics (SPH) collision simulations covering a large range of stellar masses ($0.1-75$ M$_{\odot}$) and collision velocities.\\
\indent In this work, we study a smaller parameter space than that covered by \cite{2005MNRAS.358.1133F}, but our SPH calculations are at higher resolution (typically involving $\sim7\times10^{4}$ particles) and include the highest--resolution simulations of massive stellar encounters ever performed, using $\sim6\times10^{5}$ particles. The use of two very different resolutions allowed us to demonstrate that our conclusions, in particular the rapid mergers we observe in encounters with small periastrons, are independent of numerical effects. Our superior resolution also allowed us to examine aspects of collisions and encounters between high--mass main sequence stars that \cite{2005MNRAS.358.1133F} did not study, such as the rotation of the collision products. We also performed simulations in which the encounter periastron was larger than the sum of the stellar radii -- \cite{2005MNRAS.358.1133F} explicitly neglected such encounters. Our work is particularly relevant to determining the properties and evolution of collision products and to the runaway mergers model of \cite{2002ApJ...576..899P}.\\
\indent In Section 2, we discuss our numerical techniques and the calculations we have performed. In Section 3 we give the results of our calculations. In Section 4 we discuss the possible outcomes of encounters with the collision products described in Section 3. We present our conclusions in Section 5.\\

\section{Numerical techniques}
\begin{figure*}
\includegraphics{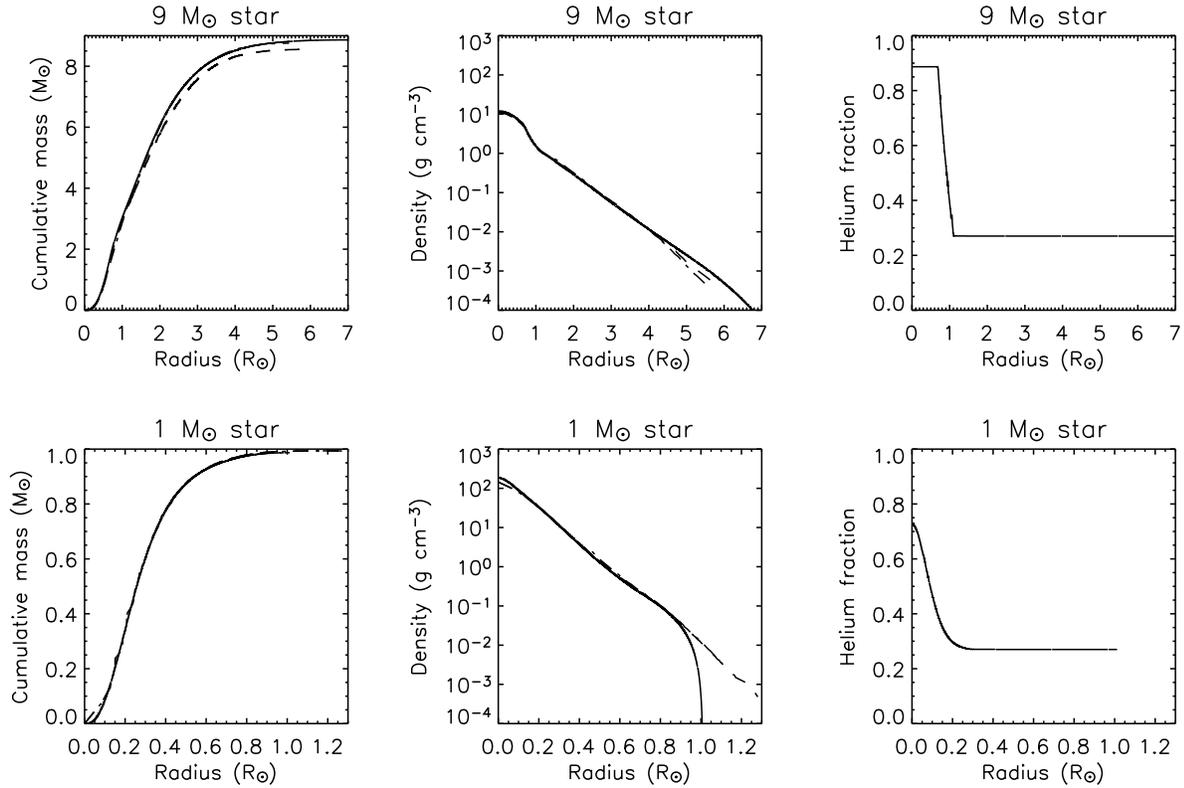}
\caption{Cumulative mass, density and composition profiles for the two main-sequence stellar models used in our calculations (dot--dashed lines: relaxed high--resolution UKAFF models, dashed lines: relaxed lower--resolution PC models, solid--lines: ATON models).}
\label{fig:init}
\end{figure*}

The SPH code used in this work is fully described in \cite{1995PhDT.......181B}. The code uses the cubic spline kernel of \cite{1985A&A...149..135M} and enforces a neighbour number of $\simeq50$ for all particles. We used the artificial viscosity prescription of \cite{1983MNRAS.204..715G}, with $\alpha=1$ and $\beta=2$. 
The majority of our SPH calculations were composed of $\approx70, 000$ particles and were run on Linux PCs. We also performed two very high--resolution calculations composed of $\approx623, 000$ particles on the SGI Origin 3800 at the United Kingdom Astrophysical Fluids Facility (UKAFF).\\
\indent We generated spherically--symmetric equilibrium stellar models using the \textsc{ATON} stellar evolution code (e.g. \cite{1998A&A...334..953V}). We extracted from these models radial profiles of density, internal energy and helium fraction and constructed three-dimensional (although initially spherically-symmetric) SPH models of the stars.\\
\indent We utilised two solar--metallicity stellar models -- a $1$ M$_{\odot}$ star approximately halfway through its main sequence life, and a $9$ M$_{\odot}$ star near the end of its main--sequence phase. We constructed our SPH stellar models by building a uniform close--packed grid of equal--mass SPH particles and iteratively adjusting the mass of each particle until the discrepancy between the local SPH density at a given radius and the corresponding density derived from the \textsc{ATON} code was less than $0.1\%$. This procedure was able to accurately model the $1$ M$_{\odot}$ star using $\geqslant11, 000$ SPH particles and the \nineg using $\geqslant 611, 000$ particles for the UKAFF calculations without generating any features pathological to the simulations. However, the density contrast between the core and the envelope of the $9$ M$_{\odot}$ star was so great that it could not be represented by a uniform grid of $\leqslant60, 000$ particles in our PC runs. We observed that the envelope of the object was poorly resolved and that the iteration procedure had generated a population of very low--mass particles in the core, a situation which can lead to unphysical results. This was a consequence of the stellar core, in which the density changes sharply over a short lengthscale, being only a few smoothing lengths across. Our solution was to build a uniform-density grid, containing a large enough space--density of SPH particles to resolve the core, out to a radius $r_{grid}<r_{max}$, where $r_{max}$ is the `real' radius of the star (see below), and to deform the grid outside the core radius $r_{core}$ (taken to be $1$ R$_{\odot}$) stretching it to a radius of $r_{max}$, leaving a uniform particle distribution inside the core while redistributing particles in the envelope to allow better resolution near the surface of the star. The deformation was radial, achieved by multiplying the $x$, $y$ and $z$ components of each particle's initial position vector $\mathbf{r}$ by a factor $f$ given by
\begin{eqnarray}
f = \left\{\begin{array}{lll}
1 & \textrm{for} &|\mathbf{r}| \leq r_{core}\\
1+\frac{|\mathbf{r}|-r_{core}}{r_{grid}-r_{core}}\left(\frac{r_{max}}{r_{grid}}-1\right) & \textrm{for} & r_{grid}>|\mathbf{r}| >r_{core}
\end{array} \right.
\label{eqn:stretch}
\end{eqnarray}
\indent Cumulative mass and density profiles for the two SPH model stars are shown in Figure \ref{fig:init} and compared to the one--dimensional models from the ATON code. Figure \ref{fig:init} also shows the helium fraction profile of the stars. The fits of the three--dimensional SPH models to the one--dimensional ATON model are clearly very good. Note that none of the $9$ M$_{\odot}$ models has a mass of exactly $9$ M$_{\odot}$. The mass of the ATON model star is $8.85$ M$_{\odot}$, that of the high--resolution SPH model is $8.75$ M$_{\odot}$ and that of the low--resolution model is $8.56$ M$_{\odot}$. The SPH models have lower masses than the ATON model because we chose to neglect material of extremely low density beyond a radius of $\approx5$ R$_{\odot}$ in the ATON model. The radius at which this cutoff occurred resulted in the UKAFF and PC models having slightly different masses. The radius of an SPH stellar model is the radius out to which the particle distribution extends plus the average smoothing length of the particles at that radius. Since the particles in the high--resolution model had smaller smoothing lengths, more particles could be fitted inside the radial cutoff, leading to a higher mass. We allowed all the SPH models to relax before beginning any simulations. Both massive SPH model stars  expanded on relaxation, the low--resolution model slightly more so because of its lower mass, leaving the massive stellar models with radii of $\approx6$ R$_{\odot}$. The convection zone delimiting the core of the \nineg is visible in Figure \ref{fig:init} as the discontinuities in the cumulative mass and density profiles at a radius of $\approx1$ R$_{\odot}$. The cumulative mass plots of the \nineg imply that the core mass of this star is $\approx2$ M$_{\odot}$ and the helium fraction plot shows that the \nineg has almost exhausted its core hydrogen. The SPH model \onesg also expanded, but the cumulative mass and density plots show that the agreement between the SPH models and the ATON models is still very good throughout almost all of the star's volume. The core of the \one, delimited by the volume of the star within which hydrogen burning is in progress, has a radius of $\approx0.2$ R$_{\odot}$ and a mass of $\approx0.3$ M$_{\odot}$. Both stars, particularly the \nine, are strongly centrally condensed and the \oneg is considerably denser throughout almost all of its volume than even the core of the \nine.\\
\indent Once relaxed, the model stars were placed on parabolic orbits with a variety of periastrons and sufficiently far apart that there was negligible tidal interaction at the beginning of each simulation.\\
\indent Encounters occurring at periastrons $r_{p}<R_{1}+R_{2}$, where $R_{1}$ and $R_{2}$ are the radii of the stars, necessarily result in a physical collision on the first periastron approach. If the two stars do not suffer such an immediate collision, they raise tides in each other and some of their orbital kinetic energy is converted into thermal and kinetic energy in the stars' envelopes. If the transfer of orbital energy to tidal energy is large enough to leave the total orbital energy negative, the stars become bound to each other. The energy required to bind the stars is dependent on their relative velocity at infinity $v_{\infty}$.\\
\indent All encounters described in this work are parabolic, so that $v_{\infty}=0$. However, from the orbital energy at the end of each of our simulations, we calculate the value of $v_{\infty}$ that \textit{would} have rendered the system just bound, which we term $\Delta v_{\infty}$. We can thus determine which encounters would result in bound systems in any cluster environment, provided that  $v_{\infty}$ is small compared with the relevant stellar escape velocities, while retaining the computational simplicity of performing parabolic calculations. This paper is concerned with stellar encounters in clusters with low velocity dispersions, $\sim10$ km s$^{-1}$. This sets the `energy scale' of our simulations.\\
\indent The physical parameters defining each of our runs are the stellar masses M$_{1}$, $M_{2}$ and the orbital periastron $r_{p}$. We list all runs performed in Table \ref{tbl:runs}.\\
\begin{table*}
\begin{tabular}{|l|l|l|l|l|l|l|l|l|}
\hline
Run no. & M$_{1}$ (M$_{\odot}$) & M$_{2}$ (M$_{\odot}$) & Periastron (R$_{\odot}$) & $r_{p}/(R_{1}+R_{2})$& No. particles & Remarks\\
\hline
1 &8.56 & 1.00 & 1.0 & 0.13 & 68, 000 &\\
2 & 8.56 & 1.00 & 2.0 & 0.26 & 68, 000 &\\
3 & 8.56 & 1.00 & 3.0 & 0.39 & 68, 000 &\\
4 & 8.56 & 1.00 & 4.0 & 0.51 & 68, 000 &\\
5 & 8.56 & 1.00 & 5.0 & 0.64 & 68, 000 &\\
6 & 8.56 & 1.00 & 7.0 & 0.81 & 68, 000 &\\
7 &8.56 & 1.00 & 10.0 & 1.30 & 68, 000 &\\
8 & 8.56 & 1.00 & 15.0 & 1.95 & 68, 000 &\\
9 & 8.56 & 1.00 & 2.0 & 0.26 & 57, 000 & as 2 using a point mass \oneg with a\\
& & & & & &  gravitational smoothing length of 0.5 R$_{\odot}$\\
10 & 8.56 &1.00 & 2.0 & 0.26 & 57, 000 & as 2 using a point mass \oneg with a\\
& & & & & & gravitational smoothing length of 0.05 R$_{\odot}$\\
11 & 8.76 & 1.00 & 1.0 & 0.13 & 623, 000 & as 2 but higher resolution\\
12 & 8.76 & 1.00 & 2.0 & 0.26 & 623, 000 & as 2 but higher resolution\\
13 & 8.56 & 8.56 & 1.0 & 0.07 & 114, 000\\
14 & 8.56 & 8.56 & 2.0 & 0.14 & 114, 000\\
15 & 8.56 & 8.56 & 3.0 & 0.21 & 114, 000\\
16 & 8.56 & 8.56 & 4.0 & 0.28 & 114, 000\\
17 & 8.56 & 8.56 & 5.0 & 0.35 & 114, 000\\
18 & 8.56 & 8.56 & 10.0 & 0.70 & 114, 000\\
19 & 8.56 & 8.56 & 15.0 & 1.05 & 114, 000\\
20 & 8.56 & 8.56 & 18.0 & 1.26 & 114, 000\\
\hline
\end{tabular}
\caption{Runs performed. Each run is defined by the masses of the stars, the periastron at which the encounter occurs and the total number of SPH particles used in the simulation.}
\label{tbl:runs}
\end{table*}

\section{Results}
\begin{figure}
\includegraphics[width=0.5\textwidth]{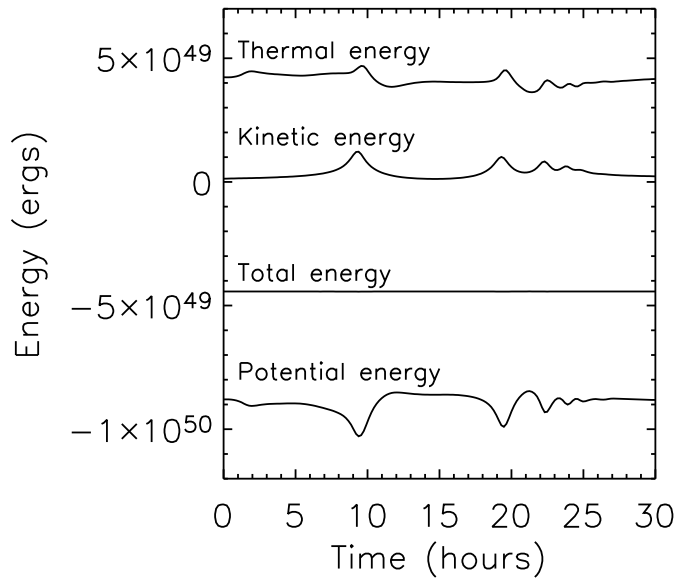}
\caption{Evolution of system energy components during the
  high--resolution encounter between a \nineg and a \oneg with a
  periastron of $2$ R$_{\odot}$. The corresponding plots for the the
  low--resolution calculation are not shown because they are difficult to distinguish.}
\label{fig:9102_energy}
\end{figure}

\begin{figure}
\includegraphics[width=0.5\textwidth]{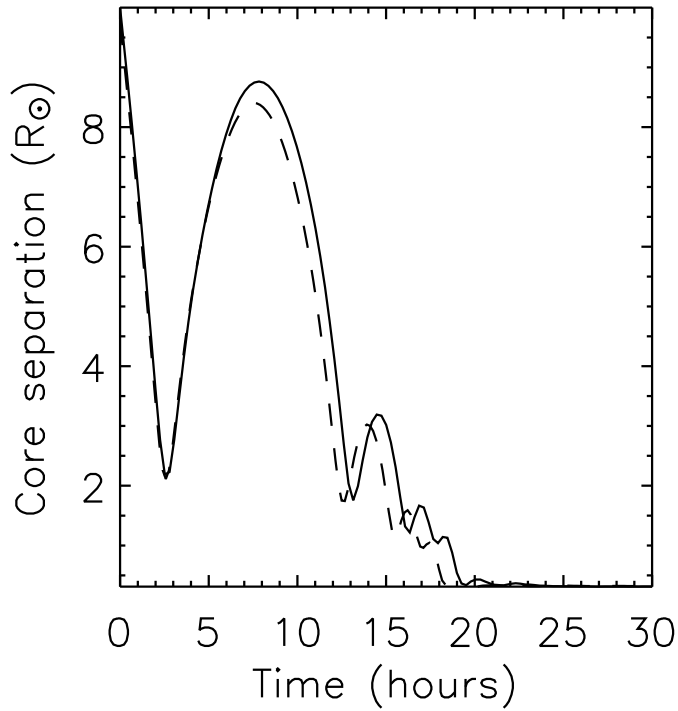}
\caption{Separation of stellar cores during the high--resolution
  (solid line) and low--resolution (dashed line) encounter between a \nineg and a \oneg with a periastron of $2$ R$_{\odot}$.}
\label{fig:9102_sep}
\end{figure}

\indent In Figure \ref{fig:9102_energy}, we plot the evolution of the
thermal, kinetic, gravitational potential and total energies during
the high-resolution UKAFF calculation involving a \nineg and a \oneg
with a periastron of $2$ R$_{\odot}$. In Figure \ref{fig:9102_sep}, we
plot the separation of the stellar cores (taken to be equal to the
separation of the stars' centres of mass) against time for the same
simulation and compare it with the corresponding PC run. We take this as our `fiducial run', against which we compare all other runs.\\
\indent There are $4$ distinct peaks visible in the energy and core separation plots corresponding to 4 periastron passages occurring before the system formed a single merged object.\\
\begin{figure}
\includegraphics[width=0.5\textwidth]{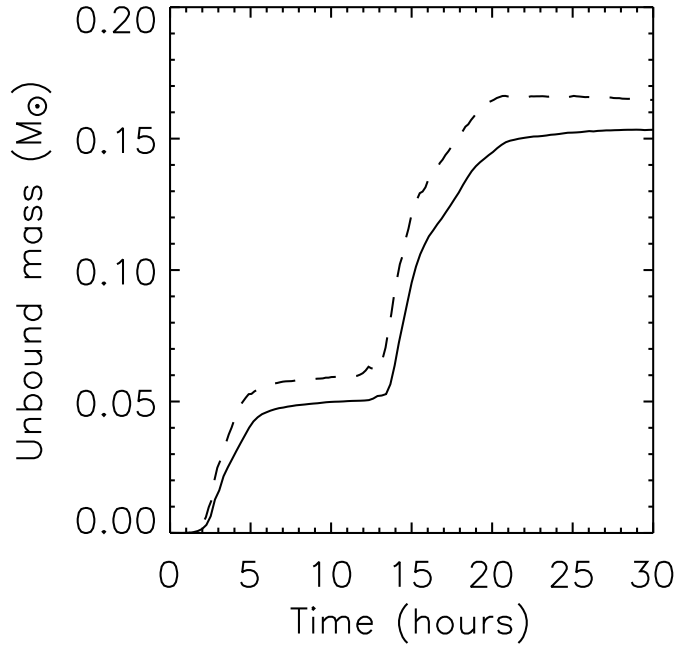}
\caption{Mass loss during the high--resolution (solid line) and
  low--resolution (dashed line) encounters between a \nineg and a \oneg with a periastron of $2$ R$_{\odot}$.}
\label{fig:9102_ml}
\end{figure}
\indent In Figure \ref{fig:9102_ml}, we follow the mass--loss in this calculation. We assume that any SPH particles with positive total energy and moving away from the centre of mass will be lost from the system. Figure \ref{fig:9102_ml} shows that mass ejection in this simulation proceeds in several distinct phases, but that the total quantity of mass lost by the time the cores merge is not large - a few tenths of a solar mass or a few percent of the combined stellar mass. Figures \ref{fig:9102_sep} and \ref{fig:9102_ml} show that the mass--loss rate increases sharply as the intruder star approaches its first periastron and decreases after the first periastron approach. The rate falls almost to zero at a time of $\approx10$ hr because, as Figure \ref{fig:9102_sep} shows, the \oneg leaves the \nine's envelope. At a time of $\approx12$ hr, the intruder falls back into the larger star and mass ejection begins again. The \oneg has now lost enough kinetic energy that it is unable to leave the envelope of the \nineg again and spirals rapidly towards the larger star's core. The cores merge at $\approx18$ hr, when the core separation becomes effectively zero. Mass loss ceases rapidly after this point.\\
\indent Quantitative differences between the UKAFF and PC runs were very small. The slightly greater mass of the \nineg in the UKAFF runs led to slightly tighter orbits, greater dissipation of kinetic energy (by $\approx1\%$) and to mass loss smaller by a few hundredths of a solar mass. We conclude from the concordance of the results that the high--mass stars studied in this paper can be accurately modelled using $\sim6\times10^{4}$ particles.\\
\subsection{Capture radius}
\begin{figure}
\includegraphics[width=0.5\textwidth]{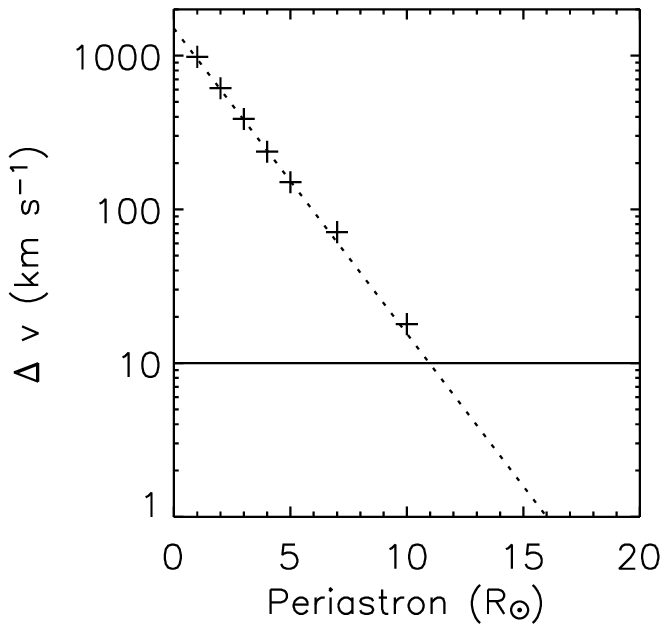}
\caption{Plot of $\Delta v_{\infty}$ after first periastron passage against initial periastron distance for encounters between \ninesg and \ones. The dotted line is a fit to the measured values and the solid line represents $\Delta v_{\infty}=10$ km s$^{-1}$. The error on each point is $\pm8$ km s$^{-1}$.}
\label{fig:91_deltav}
\end{figure}

\begin{figure}
\includegraphics[width=0.5\textwidth]{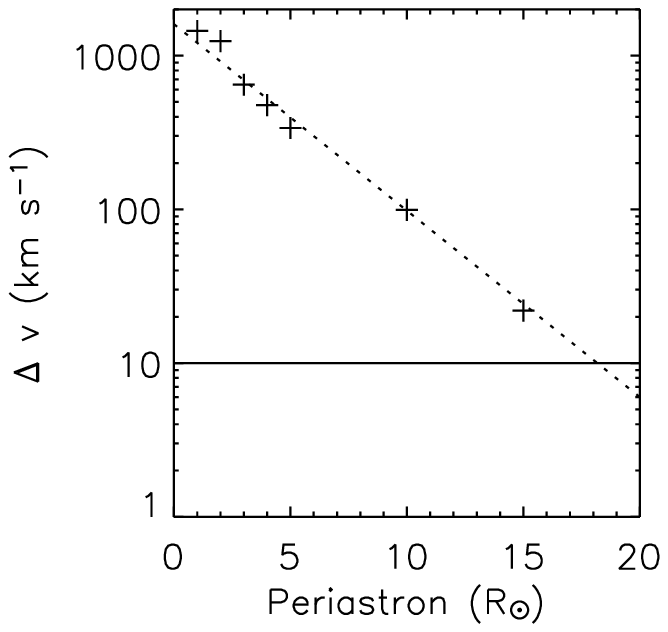}
\caption{Plot of $\Delta v_{\infty}$ after first periastron passage against initial periastron distance for encounters between two \nines. The dotted line is a fit to the measured values and the solid line represents $\Delta v_{\infty}=10$ km s$^{-1}$. The error on each point is $\pm17$ km s$^{-1}$.}
\label{fig:99_deltav}
\end{figure}

\indent To calculate the distance within which the stars must approach to become bound to each other for a given value of $v_{\infty}$, we determined the orbital parameters of each simulation after the first periastron passage. We used an iterative procedure to decide which (if either) star each SPH particle was bound to and calculated the centre--of--mass positions and velocities of both objects. From these, we determined the total orbital energy and hence $\Delta v_{\infty}$ for each initial encounter. In Figures \ref{fig:91_deltav} and \ref{fig:99_deltav}, we plot $\Delta v_{\infty}$ as a function of periastron for encounters between \ninesg and \onesg and encounters between two \ninesg respectively. Note that the last point on each plot should be treated with some caution, since the error in the value (determined by the accuracy of the SPH code's energy conservation) is comparable to the measurement. The error in the code's energy conservation is fractional -- in both cases we have multiplied the fractional error by the total energy of each run (which is different for the two types of encounter considered) and converted the result into a velocity. We are interested in stellar systems with low velocity dispersions such as the Trapezium cluster, so we define the \textit{capture radius}, $r_{c}$, to be the periastron at which $\Delta v_{\infty}=10$ km s$^{-1}$. Note that we have here implicitly assumed that $\Delta v_{\infty}$ is independent of $v_{\infty}$. We repeated the encounter between two \ninesg with a periastron of $15R_{\odot}$ on three hyperbolic orbits with $v_{\infty}=10$, $20$, and $50$ km s$^{-1}$ and found that this assumption is justified within this range of velocities, which covers the stellar systems in which we are interested.\\
\indent Figures \ref{fig:91_deltav} and \ref{fig:99_deltav} show that the capture radius for encounters between \ninesg and \onesg is $\approx11$ R$_{\odot}$ and that for the encounters between two \ninesg is $\approx18$ R$_{\odot}$. The radius of the \nineg is $\approx6$ R$_{\odot}$, implying that encounters between \ninesg and \onesg with periastrons between $7$ and $11$ R$_{\odot}$ and  encounters between two \ninesg with periastrons between $12$ and $18$ R$_{\odot}$ lead initially to a bound system of two stars. The borderline between what is considered a `captured system' and an `immediate merger' is somewhat arbitrary. For example, encounters between \ninesg and \onesg with initial periastrons of $4$ and $5$ R$_{\odot}$, which are strictly grazing collisions, produced captured systems in which the intruder stars were able to exit the envelopes of the \ninesg on very eccentric elliptical orbits. Such systems cannot be studied further by means of SPH simulations because the CPU time required is prohibitively large. However, we can study their evolution indirectly, as described in the next section.\\

\subsection{Evolution of grazing encounters}
\begin{figure*}
\includegraphics[width=1.0\textwidth]{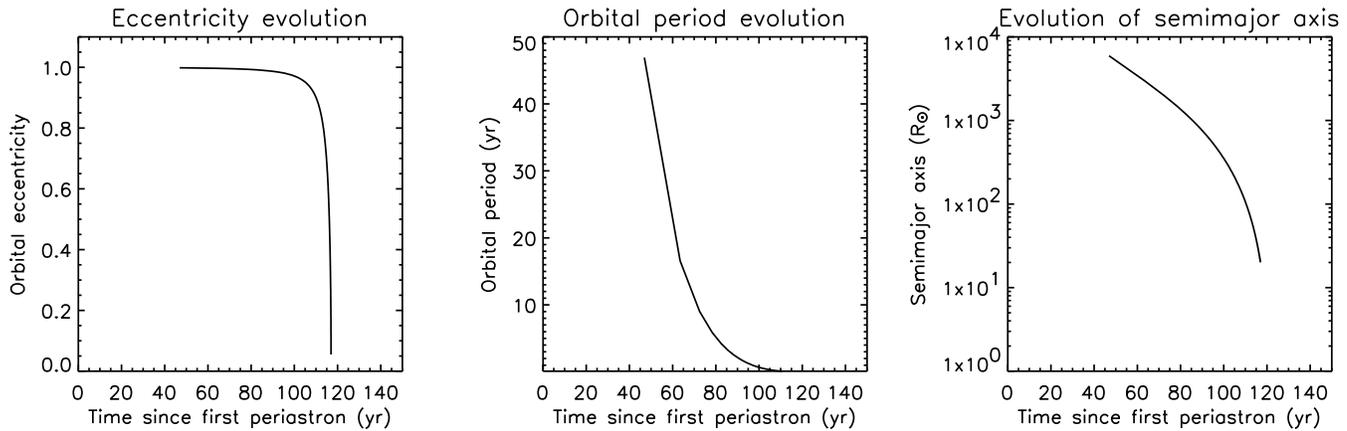}
\caption{Evolution of the orbital eccentricity, period and semimajor axis in the captured system formed by the encounter of a \nineg and a \oneg with a periastron of $10$ R$_{\odot}$.}
\label{fig:9110_oe}
\end{figure*}
In encounters where the intruder star passes within the capture radius but suffers only a  grazing collision or misses the other star completely, a captured system consisting of two stars in highly eccentric orbits results. Although we cannot study the evolution of these systems hydrodynamically, we can attempt to infer their evolution from the orbits in which they are left by their first encounter.\\
\indent We first assume that, since the periastron changes very little in tidal encounters, the quantities of tidal energy $\Delta E_{p}$ and angular momentum $\Delta J_{p}$ extracted from the orbit on the first periastron passage are also extracted at each subsequent periastron passage, and that the tidal energy taken up by the stars themselves is efficiently radiated away. For the $n$th periastron passage, we therefore have, for the energy $E_{n}$ and orbital angular momentum $J_{n}$
\begin{eqnarray}
E_{n}=E_{initial}-n\Delta E_{p}=-n\Delta E_{p},
\end{eqnarray}
where $E_{initial}$ is the initial orbital energy and is zero for parabolic encounters, and
\begin{eqnarray}
J_{n}=J_{initial}-n\Delta J_{p},
\end{eqnarray}
where $J_{initial}$ is the initial orbital angular momentum.
\indent We then calculate the eccentricity $\epsilon_{n}$, semi-major axis $a_{n}$, apastron $r_{n}$ and period $T_{n}$ for each orbit using
\begin{eqnarray}
\epsilon_{n}=\sqrt{1+\frac{2E_{n}J_{n}^{2}}{G\mu^{3}M^{2}}},
\end{eqnarray}
where $M=M_{1}+M_{2}$ and $\mu=M_{1}M_{2}/M$,
\begin{eqnarray}
a_{n}=-\frac{GM_{1}M_{2}}{2E_{n}},
\end{eqnarray}
\begin{eqnarray}
r_{n}=(1+\epsilon_{n})a_{n},
\end{eqnarray}
and
\begin{eqnarray}
T_{n}=\sqrt{\frac{4\pi^{2}a_{n}^{3}}{GM}}.
\end{eqnarray}
\indent We can then evolve the binary system according to our assumptions. The results of applying this procedure to the encounter between a \nineg and a \oneg at a periastron of $10$ R$_{\odot}$ are shown in Figure \ref{fig:9110_oe}. The effect of the tidal interaction between the stars is to circularise their orbit, so the eccentricity, semimajor axis and orbital period decrease rapidly until, after $\approx115$ yr, the apastron is approximately equal to the periastron and the peraistron itself begins to decrease. At this point we assume that the system will rapidly merge.\\
\indent We determined the binary grind-down times for all of our encounters and found that they ranged from a few hundred hours for encounters with periastrons of $4$ R$_{\odot}$ to $\approx100$ yr for  encounters between \ninesg and \onesg with periastrons of $10$ R$_{\odot}$ or encounters between two \ninesg with periastrons of $15$ R$_{\odot}$.\\
\indent \cite{1995ApJ...450..722M} and \cite{1995ApJ...450..732M} and
also \cite{1992ApJ...385..604K} studied the evolution of tidal capture
binaries using numerical schemes in which the cores of the stars were
represented by point masses and the polytropic envelope of the larger
star was decomposed into spherical harmonics. The point masses and
envelope components were allowed to exchange energy under the
assumption that the overall interaction was adiabatic. This allowed
the authors to integrate systems similar to those formed in our calculations for many thousands of orbits. Mardling found that, for very eccentric orbits (as the orbits of our captured systems are, since they were captured from parabolae), the evolution of the binary separation was unpredictable but that, in general, very eccentric binaries did not grind down to mergers. \cite{1992ApJ...385..604K} found that binaries formed by encounters at a few stellar radii were often scattered into wide orbits on subsequent periastron passages and that on, average, such binaries spent most of their time as soft binaries and were therefore vulnerable to disruption.\\
\indent There is a third, more likely outcome of tidal capture. If the binary grind--down times $t_{grind}$ and final separations $a_{f}$ estimated by our method above are realistic, the tidal power $L_{tidal}$ deposited in the envelopes may be estimated from 
\begin{eqnarray}
L_{tidal}\sim\frac{GM_{1}M_{2}}{2a_{f}t_{grind}}.
\end{eqnarray}
The tidal luminosity of the encounter between a \nineg and a \oneg is
then $\sim 10^{5}$ L$_{\odot}$, far in excess of the luminosity of a
main--sequence \nine. It is therefore likely that the tidal interaction will cause the star(s) to expand and enter a phase of common--envelope evolution.\\

\subsection{Common envelope evolution}
\indent If our captured systems do grind down tidally, the likely result is a 
common--envelope system in which the stellar cores inspiral inside a common envelope derived from the stars' original envelopes. The fate of a common envelope system hinges on whether the energy released during the inspiral is sufficient to unbind the envelope before the cores merge.\\
\indent If the common--envelope phase is assumed to expel the envelope and if the energy released derives purely from the initial orbital energy of the inspiralling cores, one may derive an expression for the final separation of the cores, $a_{f}$ as a function of their initial separation $a_{i}$ (e.g. \cite{2000A&A...360.1043D}),
\begin{eqnarray}
a_{f}=M_{i}M_{c}\left(\frac{2(M_{c}+M_{env})M_{env}}{\alpha\lambda R}+\frac{M_{i}(M_{c}+M_{env})}{a_{i}}\right)^{-1}.
\label{eqn:ce}
\end{eqnarray}
where $M_{i}$ is the mass of the intruder star, $M_{c}$ is the core
mass of the larger star, $M_{env}$ is the mass of the larger star's
envelope, $\alpha$ is a parameter expressing the efficiency with which
the orbital energy of the cores is used to unbind the envelope and
$\lambda$ is a geometrical factor accounting for the envelope's
structure. Formally, $\alpha$ may take any positive value; $\alpha=1$
implies that all the orbital binding energy is required to eject the
common envelope, and $\alpha>1$ implies that some additional source of
energy is required. To determine the outcome of common--envelope
evolution for a given system, the product $\alpha\lambda$ can be
treated as a parameter.\\
\indent If the separation between the cores becomes so small that one
of them fills its Roche lobe (note that, since these systems are not
corotating, the use of Roche geometry is an approximation, but it
proves to be adequate), the cores merge and energy deposition in the envelope ceases.  For two cores $i$ and $j$ of mass $m_{i}$ and $m_{j}$ at a separation $a$, the size of the Roche lobe of core $i$ is given by \citep{1983ApJ...268..368E}
\begin{eqnarray}
r_{R}^{i}=\frac{0.49(m_{i}/m_{j})^{\frac{2}{3}}a}{0.6(m_{i}/m_{j})^{\frac{2}{3}}+ln(1+(m_{i}/m_{j})^{\frac{1}{3}})}.
\label{eqn:rlof}
\end{eqnarray}
\indent We continuously estimated the
Roche lobe size for both cores from their masses and separation and
using Equation \ref{eqn:rlof} as the
simulation evolved. We compared the estimated Roche lobe size of each
core to that core's physical size. The cores of the initial model
stars were defined by the hydrogen--burning regions in the models'
centres and all SPH particles within this region were tagged at the
start of each simulation. We then estimated the sizes of the stellar
cores from the positions of all the core particles, assuming that each
core neither lost nor gained particles and that the cores were
spherical. At the point in time where the radius of either core calculated in
this way first exceeded the Roche lobe size of that core, an animation
of the simulation was examined to confirm that mass transfer between
the cores was in progress. In both cases, we used the encounters with a periastron of $2$ R$_{\odot}$. In encounters between \ninesg and \ones, core Roche--lobe overflow occurs at a separation of $\approx1.2$ R$_{\odot}$ and in encounters between two \nines, it occurs at a separation of $\approx1.7$ R$_{\odot}$. Taking the initial separations for both types of encounter to be the relevant capture radii (as would be the case following the tidal grind--down of a captured system), we may now use Equation \ref{eqn:ce} to determine the minimum value of $\alpha\lambda$ that would result in each system ejecting its envelope before undergoing core merger.\\
\indent For encounters between \ninesg and \ones, we took the core,
envelope and intruder masses to be $2$ M$_{\odot}$, $7$ M$_{\odot}$
and $1$ M$_{\odot}$ respectively. We found that we would require
extreme values of $\alpha\lambda$ ($>20$) to eject the envelope before
core merger occurs in the case of an inspiral from the capture radius
for these encounters, $11$ R$_{\odot}$. We repeated the analysis for
encounters between two \nines, taking the core, envelope and intruder
masses to be $2$ M$_{\odot}$, $14$ M$_{\odot}$ and $2$ M$_{\odot}$
respectively. In the case of inspiral from a radius of $18$
R$_{\odot}$ (the capture radius for these encounters), ejection of the
envelope requires even more extreme values of $\alpha\lambda$,
$>100$. The value of $\lambda$ for the \textit{initial} \nineg models was
$\simeq7$. This value can be expected to increase as the \nineg
expands, but we would probably still require values of $\alpha$ in
excess of unity to expel the envelope, implying that some source of
energy other than the orbital energy of the inspiralling cores would
be required. Such a scenario is very difficult to study in the energy--conserving model of common--envelope evolution, but another prescription for studying common--envelope systems was suggested by \cite{2004RMxAC..20...39N}\\
\indent \cite{2004RMxAC..20...39N} and  \cite{2005MNRAS.356..753N} developed an approach to common envelope evolution based on explicit conservation of angular momentum. They assume that the fractional quantity of angular momentum carried away by ejection of the envelope is proportional to the fractional mass loss, so that
\begin{eqnarray}
\frac{\Delta J}{J}=\gamma\frac{\Delta M}{M}=\gamma\frac{M_{env}}{M_{c}+M_{env}}.
\label{eqn:n_and_t}
\end{eqnarray}
One may then derive an expression for $a_{f}$ as a function of $a_{i}$ analogous to Equation \ref{eqn:ce}:
\begin{eqnarray}
a_{f}=a_{i}\left(\frac{(M_{env}+M_{c})M_{i}}{M_{c}M_{i}}\right)^{2}\left(\frac{M_{c}+M_{i}}{M_{env}+M_{c}+M_{i}}\right)\times\nonumber\\
\left(1-\gamma\frac{M_{env}}{M_{env}+M_{c}+M_{i}}\right)^{2}.
\label{eqn:ce2}
\end{eqnarray}
\indent In their study of white dwarf binaries as the products of common envelope evolution, \cite{2005MNRAS.356..753N} concluded that the origins of all the systems they examined could be explained using the above expressions if $1.5<\gamma<1.75$. We applied their analysis to our stellar encounters, again assuming that the grind--down of a captured binary resulted in common--envelope evolution where the initial separation between the two cores corresponded to the appropriate capture radius. We found that the analysis of Nelemans and Tout (2004) predicts that, under these conditions,  encounters between \ninesg and \onesg will result in expulsion of the envelope for $\gamma\gtrsim1.65$ and that encounters between two \ninesg will result in envelope expulsion for any value of $\gamma$ in Nelemans and Tout's preferred range. In both cases, the final separations of the systems were very sensitive to the chosen value of $\gamma$.\\
\indent The results of the calculations performed in this section are inconclusive. We have applied two different common--envelope formalisms to encounters between \ninesg and \onesg and between two \ninesg and found that the traditional explicitly energy--conserving formalism implies that envelope ejection is unlikely and that these encounters probably result in mergers, whereas the explicitly angular--momentum--conserving formalism of \cite{2005MNRAS.356..753N} implies that envelope ejection is possible in both cases and that the stellar cores will not merge. This essentially means that the orbits of these common--envelope systems have sufficient angular momentum, but insufficient energy, to expel their envelopes. Settling the question of whether or not the systems merge rests on whether some additional source of energy can be tapped, so that the envelope can be expelled while conserving energy \textit{and} angular momentum.  \cite{2004NewA....9..399S} suggests that angular momentum deposition causes the envelope to swell and triggering the formation of dust. The resulting dust--driven wind taps the stellar luminosity, which provides the `extra' source of energy required to eject the envelope. In the absence of an additional energy source, the result of common--envelope evolution would presumably be that a small quantity of material would be expelled from the systems, carrying all the orbital angular momentum, although it is not obvious how this might come about.\\
\indent We have only considered massive stars of a single mass. On the
main sequence, the fraction of the stellar mass made up by the core is
an increasing function of stellar mass. We repeated our common
envelope analysis for an encounter between a $50$ M$_{\odot}$ star,
with a $35$ M$_{\odot}$ core and a $15$ M$_{\odot}$ envelope
\citep{1994sse..book.....K}, and a $1$ M$_{\odot}$ intruder. We found
that a merger could be avoided and the stellar envelope ejected in
common--envelope evolution following binary grind--down for
$\alpha\lambda\gtrsim5$ and for any value of $\gamma$ in the range
preferred by Nelemans and Tout. This result is counterintuitive at
first sight, but follows from the fact that the core of the $50$
M$_{\odot}$ star has such a large mass, so that there is a large
quantity of orbital energy available with which to expel the envelope,
and also that the envelope has a relatively small mass and binding
energy. Similar conclusions have been reached by other authors,
e.g. \cite{1994AAS...18512104T}, \cite{1995ApJ...445..367T}, who found that a $24$ M$_{\odot}$
supergiant star entering a common--envelope phase with a $1.4$
M$_{\odot}$ neutron star would lose its envelope.\\

\subsection{Encounters where $r_{P}<(R_{1}+R_{2})$}
\begin{figure}
\includegraphics[width=0.45\textwidth]{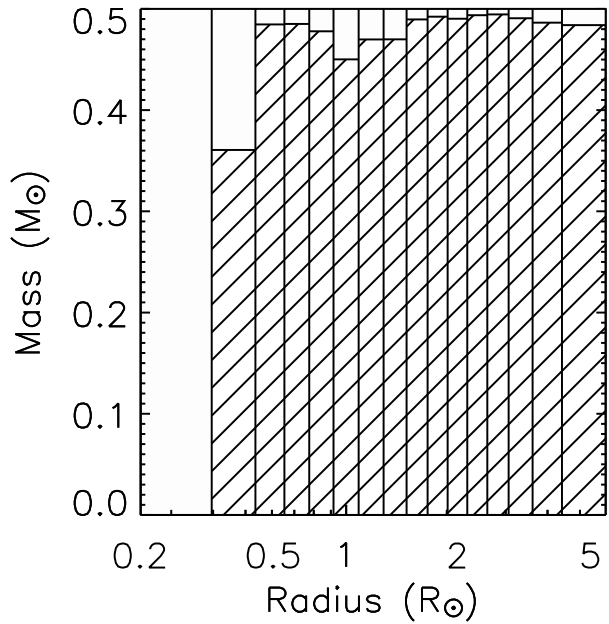}
\caption{Histogram showing to which star material in $0.5$ M$_{\odot}$ shells within the collision product of the encounter between a \nineg and a \oneg with a periastron of $2$ R$_{\odot}$ originally belonged. Hatched material originated in the \nineg and white material in the \one.}
\label{fig:91_hist}
\end{figure}

In this section, we discuss encounters in which the periastron distance on the first approach is less than the sum of the stellar radii, so that the stars collide immediately. The dissipative interaction between the two stars ejects mass from the system and the cores of the stars spiral inwards.\\
\indent We integrated the encounters between \ninesg and \onesg with
periastrons of $1$, $2$ and $3$ R$_{\odot}$ and those between two
\ninesg with periastrons of $1$, $2$, $3$ and $4$ R$_{\odot}$ until
the cores of the stars merged (doing this for runs with larger
periastrons was not possible, since the stars were captured onto
relatively wide orbits which would require prohibitive quantities of
CPU time to integrate). In all the simulations which we ran to core
merger, the mergers were very rapid ($\approx15-80$ hours in the
encounters between \ninesg and \onesg and $\approx12-150$ hours in the
encounters between two \nines). The total quantities of mass ejected
by the mergers were small in comparison to the sum of the progenitor
masses ($\approx0.1$ M$_{\odot}-0.3$ M$_{\odot}$ in encounters between
\ninesg and \onesg and $\approx0.4$ M$_{\odot} - 0.9$ M$_{\odot}$ in
the encounters between two \nines). These results are largely
consistent with our common--envelope calculations in the sense that
the calculations presented in Section 3.3 suggest that encounters
between \ninesg and \onesg with these periastrons should not expel the
envelope of the \nineg for any value of $\alpha\lambda$ or for any
value of $\gamma$ considered reasonable by
\cite{2004RMxAC..20...39N}. Envelope ejection in encounters between
two \ninesg with periastrons $\leqslant4$ R$_{\odot}$ is forbidden by
the energy--conserving analysis, although the $\gamma$-formalism
suggests that envelope ejection may occur for values of
$\gamma\gtrsim1.6$. Note, however, that we are not here suggesting
that common--envelope evolution is generally amenable to study by SPH
simulations, since integration for many dynamical times would be
required in the majority of cases. Apart from the prohibitively long CPU times required, a more accurate artificial viscosity prescription would probably be required, since significant momentum transport by viscosity may occur over a large number of orbits. In our SPH calculations, mergers occur in only a few tens of dynamical times and almost all momentum transfer or transport is by gravitational torques or the ram--pressure interaction between the stars. It would also be difficult to include other possible sources of energy, such as dust--driven winds. Our SPH calculations implicitly assume that no such sources of energy are available and hence that $\alpha$ does not exceed unity.\\
\indent To determine the probable subsequent evolution of the products of these mergers, we examined the distribution of the material from the progenitor stars in the collision products. In Figure \ref{fig:91_hist}, we divide the collision product of the encounter between a \nineg and a \oneg with a periastron of $2$ R$_{\odot}$ into spherical $0.5$ M$_{\odot}$ shells and show what fraction of the material in each shell originated from the \oneg and what fraction came from the \nine.\\
\indent The \oneg has survived the merger largely intact and occupies
the centre of the merger product. Since the \oneg is considerably
denser and cooler than the \nine, it is not surprising that it has
sunk in this way. In the corresponding encounter between two \nines,
we observed that the collision product was well mixed after $\approx
30$ hours. We conclude, in common with \cite{2002ApJ...568..939L},
\cite{1997ApJ...477..335S} and \cite{2002MNRAS.332...49S} (and
contrary to \cite{1987ApJ...323..614B}) that little
\textit{hydrodynamic} mixing occurs during collisions of main sequence
stars of unequal mass, although it is possible that mixing will occur
when the merger product is allowed to approach thermal equilibrium,
since the core will then become convective.\\
\begin{figure}
\includegraphics[width=0.45\textwidth]{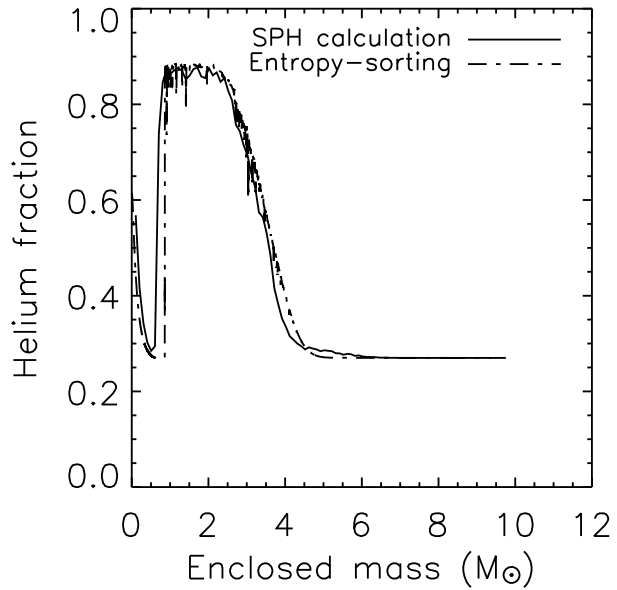}
\caption{Plot of the helium fraction by mass against enclosed mass
  within the collision product of the encounter between a \nineg and a
  \oneg with a periastron of $2$ R$_{\odot}$. We compare the results
  of the SPH calculation with those derived using the entropy--sorting
  method of \protect\cite{2002ApJ...568..939L}. The dashed plot was
  generated by inserting our initial stellar models into the Make Me A Star package available at \texttt{http://faculty.vassar.edu/lombardi/mmas/}.}
\label{fig:91_mix}
\end{figure}
\indent In Figure \ref{fig:91_mix}, we plot the fraction by mass of
helium against enclosed mass in the collision product and compare the
results from our SPH calculation from results derived using the
entropy--sorting technique suggested by
\cite{2002ApJ...568..939L}. The two methods evidently agree closely,
both suggesting that the merger remnant will have a core with a
helium--rich centre surrounded by a layer rich in fresh hydrogen, in
turn surrounded by a layer enriched in helium. If the merger remnant
eventually comes to approximate a normal main sequence $10$
M$_{\odot}$ star, it will have a core mass of $\approx3$
M$_{\odot}$. The combined core masses of the progenitor stars is
$\approx2.3$ M$_{\odot}$, so the hydrogen--burning region will need to
grow before the remnant settles onto the main sequence. The innermost
$3$ M$_{\odot}$ of the merger remnant has a mean helium fraction of
$\approx70\%$, whereas the core of the original \nineg had a core
helium fraction of $\approx90\%$. The duration of the main--sequence
phase for the model $9$ M$_{\odot}$ star generated using the ATON code
was $3.0\times10^{7}$ yr. We used the model at an age of
$2.8\times10^{7}$ yr in our SPH calculations. If the merger product
eventually settles back onto the main sequence, it should continue to
burn hydrogen for $\approx6\times10^{6}$ yr, whereas the original
\nineg had enough hydrogen left to burn for only another
$\approx2\times10^{6}$ yr. The merger may then extend the
hydrogen--burning lifetime of the \nineg by $\approx4\times10^{6}$ yr,
$\approx13\%$ of its total main--sequence lifetime. This is an extreme
case of rejuvenation. If the $9$ M$_{\odot}$ star were younger and
thus had a less helium--enriched core, the effect of the merger on the
core helium fraction would clearly be smaller. The merger of two
evolved stars, as in our mergers of two \nines, also clearly does not
result in a remnant with a much extended hydrogen--burning phase,
since the remnant core is formed from two already highly
helium--enriched cores.\\

\subsection{Massive stars with rapidly--rotating cores - hypernova progenitors?}
\indent Hypernovae (HNe) are a class of unusually luminous supernovae, specifically Type Ic events with luminosities $>$few$\times10^{51}$ erg \citep{1999AN....320..265N}. Recently, a connection between HNe and long--duration Gamma Ray Bursts (GRBs) (all references to GRBs in this paper relate to \textit{long--period} bursts) has been posited (see, e.g. \cite{2004ApJ...607L..17P}, suggesting that they may be different facets of the same type of event, which is itself a high--energy subset of Type Ic supernovae. 
HNe/GRB models involving stellar progenitors require extreme rotation rates to ensure that the collapsing stellar core forms a disk around its central black hole. Although young massive stars are often rapidly rotating, there are numerous means by which they can lose their angular momentum during their main--sequence lifetimes (e.g. via winds).\\
\indent Mergers are an ideal candidate for producing rapidly--rotating
massive stars, since all the orbital angular momentum of the system must reside either in the merger product or in the small amount of mass ejected in the collision. Mergers in which little mass is lost therefore result in remnants with very large angular momenta.\\
\indent For the core of a massive star to form a HN/GRB, the specific
angular momentum of material at the edge of the core before collapse begins must exceed the specific angular
momentum of material on the last stable orbit around the black hole,
so that the whole core cannot collapse into the hole, giving $j>\sqrt{6}GM_{core}/c\approx2\times10^{16}(M_{core}/2M_{\odot})$ erg s g$^{-1}$\citep{2004ApJ...607L..17P}. The merger of two \ninesg may therefore result in a hypernova progenitor if the core of the product is rotating sufficiently rapidly.\\
\indent We examined the remnant of the encounter between two \ninesg with a periastron of $2$ R$_{\odot}$ to see if it fulfils \citep{2004ApJ...607L..17P}'s criterion. Defining the core radius as the radius containing $2$ M$_{\odot}$, the core radius of the remnant is $\approx 0.8$ R$_{\odot}$ and the specific angular momentum at this radius is $\sim$ fifty times that required to prevent the core of the collision product collapsing directly into a central black hole. We were only able to analyse the collision remnant $\approx100$ hr after impact and angular momentum transport is still underway within the remnant, largely by means of spiral density inhomogeneities. However, the remnant can lose $98\%$ of its angular momentum and still satisfy \citep{2004ApJ...607L..17P}'s criterion.\\

\section{Discussion}
\subsection{Comparison with runs using point mass impactors}
\begin{figure}
\includegraphics[width=0.45\textwidth]{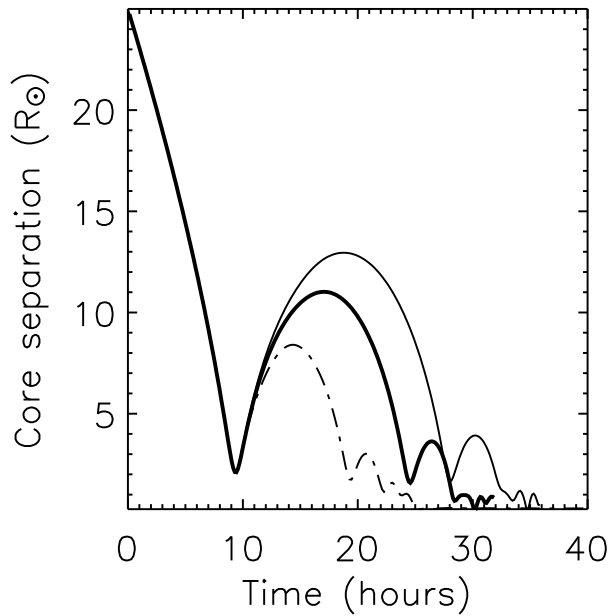}
\caption{Comparison of the evolution of the core separations with time in the pure--SPH and point mass encounters between \ninesg and \onesg with periastrons of $2$ R$_{\odot}$. Dot--dashed line: pure--SPH calculation, thick solid line: point mass calculation with gravitational smoothing length of $0.05$ R$_{\odot}$, thin solid line: point mass calculation with gravitational smoothing length of $0.5$ R$_{\odot}$.}
\label{fig:pm_core_sep}
\end{figure}

\begin{figure}
\includegraphics[width=0.45\textwidth]{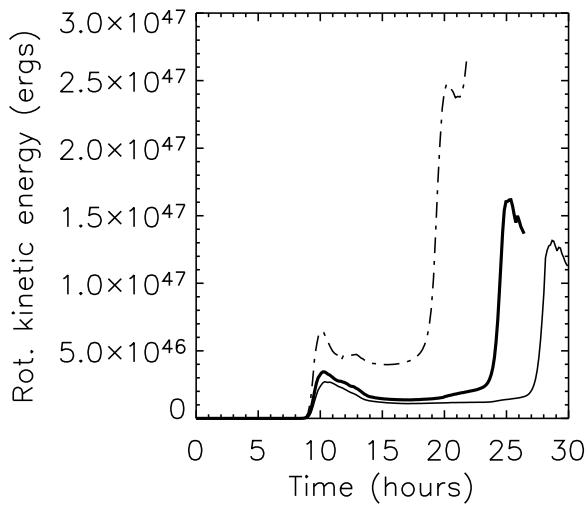}
\caption{Evolution of the rotational kinetic energy of the \nineg during the first few periastron passes in the pure--SPH and point mass encounters between \ninesg and \onesg with a periastron of $2$ R$_{\odot}$. Dot--dashed line: pure--SPH calculation, thick solid line: point mass calculation with gravitational smoothing length of $0.05$ R$_{\odot}$, thin solid line: point mass calculation with gravitational smoothing length of $0.5$ R$_{\odot}$.}
\label{fig:pm_rke}
\end{figure}

\begin{figure}
\includegraphics[width=0.45\textwidth]{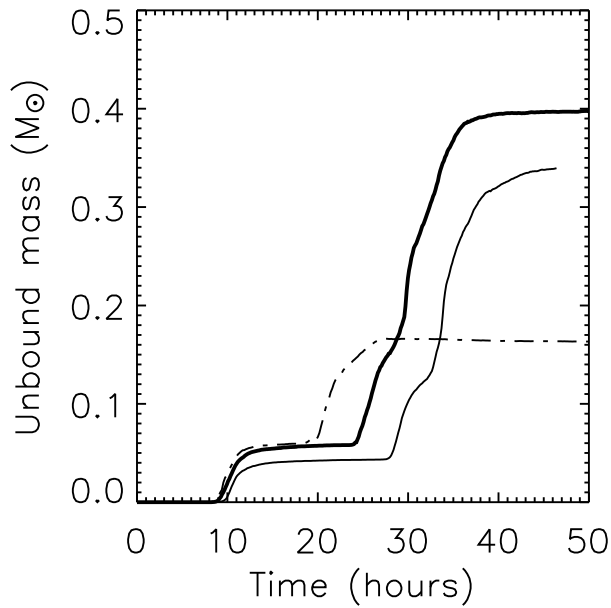}
\caption{Comparison of the evolution of the mass loss in the pure--SPH and point mass encounters between \ninesg and \onesg with a periastron of $2$ R$_{\odot}$. Dot--dashed line: pure--SPH calculation, thick solid line: point mass calculation with gravitational smoothing length of $0.05$ R$_{\odot}$, thin solid line: point mass calculation with gravitational smoothing length of $0.5$ R$_{\odot}$.}
\label{fig:pm_ml}
\end{figure}

\indent As well as the high--resolution calculations described in Section 3, we conducted two  simulations of encounters between \ninesg and \onesg with a periastron of $2$ R$_{\odot}$ in which we represented the \oneg by a point mass with gravitational smoothing lengths of $0.5$ and $0.05$ R$_{\odot}$, chosen so that the point mass smoothing length would be comparable to the SPH smoothing lengths in the \nine's envelope and core respectively. We compare these runs with the pure--SPH run with a periastron of $2$ R$_{\odot}$.\\
\indent The main difference between the pure--SPH and point--mass runs was that there was greater dissipation in the pure--SPH runs. The SPH intruder star experiences ram pressure, which the point mass does not, and thus loses kinetic energy more rapidly during its passage through the \nine. This phenomenon is discussed by \cite{1992ApJ...389..546B}, who also performed collision simulations using point--mass and SPH impactors, and can be seen clearly in Figure \ref{fig:pm_core_sep}, which contrasts the time--evolution of the core separation in the pure--SPH and point-mass runs, and in Figure \ref{fig:pm_rke} where we compare the quantities of \textit{rotational} kinetic energy transferred to the \nineg on the first few periastron passages in each run. The smaller dissipation in the point--mass simulations results in merger timescales which are longer than those in the pure--SPH calculations, but are still only of the order of tens of hours.\\
\indent The mass--loss evolution of these calculations is shown in Figure \ref{fig:pm_ml}. The mass--loss in the point--mass simulations is greater because gas particles in these runs are strongly accelerated in the deep potential wells of the point masses. A smaller gravitational smoothing length produces a deeper well and thus higher acceleration (behaviour of this kind was discussed at length by \cite{1993A&A...272..430D}). When comparing results of runs using SPH and point--mass intruders, the point masses will always eject more mass in this way than an equal--mass SPH intruder, unless their gravitational smoothing lengths are considerably larger than the SPH intruder itself.\\
\indent We conclude from these results that replacing an SPH impactor
with a point mass in stellar collision calculations produces
qualitatively similar results, but is unsuitable if one wants to study
the energetics of the encounters quantitatively. Point mass impactors
are more suitable for studies of encounters involving compact objects,
e.g. \cite{1995ApJ...445..367T}, \cite{2005ApJ...627..277L}.\\ 

\subsection{Merger products as hypernova progenitors}
We showed in Section 3 that the cores of collision remnant from encounters between two \ninesg are rotating sufficiently rapidly to fulfil the criterion outlined by \cite{2004ApJ...607L..17P}, making them plausible candidates for HNe/GRB progenitors. We now consider whether the cluster environments in which massive stars are typically found permit collision rates sufficiently large to account for a significant fraction of HNe.\\
\indent The stellar densities in globular cluster cores often reach $10^{4}$ pc$^{-3}$, although the massive stars in such systems have long since exploded. If we consider younger systems which still possess most of their high--mass stars, we find that, for example, the Orion Nebula Cluster has a stellar density of $\sim10^{4}$ pc$^{-3}$ \citep{1994AJ....108.1382M}, the putative Galactic starburst cluster NGC 3603 has a mass density of $\sim10^{5}$ pc$^{-3}$ \citep{1995A&A...300..403H} and the Arches cluster near the Galactic Centre has a mass density of $>5\times10^{5}$ pc$^{-3}$ \citep{2002ASPC..263..287F}. If we take $10^{4}$ pc$^{-3}$ as being representative of typical clusters containing massive stars, we can use Equation \ref{eqn:tenc} to estimate the fraction of massive main--sequence stars in these systems that will undergo collisions. Taking these stars to have radii of $10$ R$_{\odot}$, masses of $10$ M$_{\odot}$ and taking the cluster velocity dispersion to be $10$ km s$^{-1}$, we find that the encounter timescale is $\sim10^{10}$ yr. If the massive stars in question have main--sequence lifetimes of $\sim10^{7}$ yr and they spend their entire lives in a cluster environment, $\sim0.1\%$ of them will suffer collisions in clusters with number densities of $10^{4}$ pc$^{-3}$.\\
\indent The total rate of Type Ib/c supernovae is $\sim10^{-3}$ yr$^{-1}$ galaxy$^{-1}$ (\cite{2004ApJ...607L..17P}). Podsiadlowski et al estimate that $5\%$ of observed SNe Ib/c are HNe. Taking into account the fact that HNe are brighter than average SNe Ib/c, they estimate that the HNe rate is $\sim10^{-5}$ yr$^{-1}$ galaxy$^{-1}$. The total rate of \textit{all} core-collapse supernovae is $\sim7\times10^{-3}$ yr$^{-1}$ galaxy$^{-1}$, implying that $\sim0.1\%$ of all massive stars explode as HNe. We therefore find that the fraction of massive stars likely to suffer collisions in stellar systems with number densities typical of young massive clusters is of the same order as the fraction of massive stars that explode as hypernovae.\\
\indent To explode as a Type Ic supernova, the merger remnant must
first lose its envelope. Winds are an obvious means of achieving
this and common--envelope evolution may also expel a significant fraction of the envelope of the merger product during the merger.\\

\subsection{Ultimate fate of collision remnants}
\begin{figure*}
\includegraphics[width=0.95\textwidth]{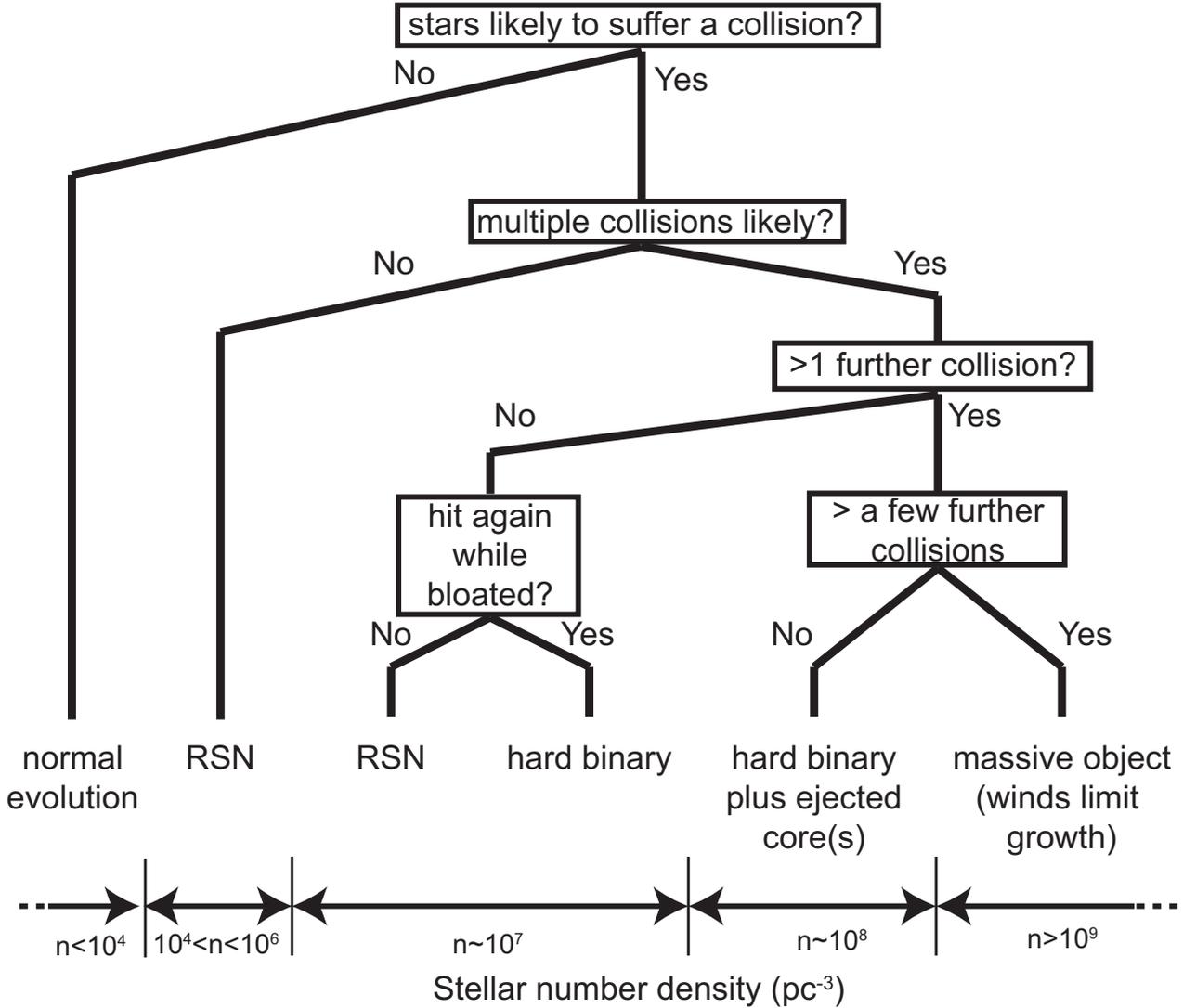}
\caption{Likely fate of the products of collisions in environments with
  the number densities given in Table \ref{tbl:numdens}. 'RSN'
  indicates that these collision remnants will explode as
  rapidly--rotating supernovae (and those with masses of $\gtrsim20$
  M$_{\odot}$ possibly as hypernovae). As explained in the text, any
  stellar cluster may pass through some or all of these density
  regimes as it evolves and may exhibit some or all regimes at any
  given moment in time.}
\label{fig:tree}
\end{figure*}

We have so far discussed only single collisions between main sequence
stars, such as occur infrequently in clusters of high stellar
density. We showed in Section 4.2 that in clusters with densities of
$\sim10^{4}$ pc$^{-3}$, $\sim0.1\%$ of massive stars are likely to
suffer collisions. This fraction increases linearly with stellar
density, reaching $100\%$ at a density of $\sim10^{7}$ pc$^{-3}$. The
extreme stellar densities ($\sim10^{8}$ pc$^{-3}$) thought to exist in
some young stellar systems \citep{1998MNRAS.298...93B} bring with them
the possibility of \textit{multiple} collisions. To understand the
fate of an encounter involving the product of an earlier merger, the
evolution of merger remnants must be considered.\\
\indent The gravitationally--focused cross section of stars of mass $M$ and radius $R$ in a stellar cluster with velocity dispersion $v_{\infty}$ is given by
\begin{eqnarray}
\sigma=\pi R^{2}\left(1+\frac{2GM}{Rv_{\infty}^{2}}\right).
\label{eqn:x-section}
\end{eqnarray}
If $v_{\infty}$ is small (a value of $10$ km s$^{-1}$ is assumed
throughout this paper), the second term inside the brackets of Equation \ref{eqn:x-section} dominates and 
an encounter timescale can be derived \citep{1975MNRAS.172P..15F}
\begin{eqnarray}
t_{enc}=7\times10^{10}\left(\frac{n}{10^{5}\textrm{pc}^{-3}}\right)^{-1}
\left(\frac{M}{\textrm{M}_{\odot}}\right)^{-1}\left(\frac{R}{\textrm{R}_{\odot}}\right)^{-1}\times \nonumber\\
\left(\frac{v_{\infty}}{10\textrm{kms}^{-1}}\right)\textrm{yr},
\label{eqn:tenc}
\end{eqnarray}
where $n$ is the stellar number density of the cluster. In this paper we have neglected the possibility that the presence of primordial binaries or the formation of tidal--capture binaries will probably decrease the effective encounter timescales, since binaries have much larger encounter cross--sections than single stars. \cite{1999A&A...348..117P} suggest that binary--formation may increase collision rates by a factor of $\sim10$.\\
\indent In the encounters between \ninesg and \ones, almost all of the kinetic energy from the \oneg was deposited in the merger product's envelope. The quantity of energy deposited in the envelope of the \nineg in the encounter with a periastron of $2$ R$_{\odot}$ was $\sim10^{7}$ L$_{\odot}$ yr. The luminosity of a $10$ M$_{\odot}$ star is $10^{3}-10^{4}$ L$_{\odot}$, so this energy will be radiated away on a timescale of $10^{3}-10^{4}$ yr. The thermal timescale of the remnant is $\sim10^{4}$ yr. The merger remnant is therefore likely to swell in a manner similar to a star leaving the main sequence and evolving into a giant. If the luminosity of the merger product remains constant, the post--main sequence track for a $10$ M$_{\odot}$ star (e.g. \cite{1994sse..book.....K}) suggests that the remnant will swell by a factor of $\sim100$ to a radius of $\sim500$ R$_{\odot}$. After $\sim10^{4}$ yr, the collision remnant will have radiated away the excess energy deposited in its envelope and will shrink back to the size of a main--sequence $10$ M$_{\odot}$ star. The collision cross section of the bloated remnant, $\sigma_{bloat}$ is $\sim100$ times that of the remnant on the main sequence, $\sigma_{MS}$ and the time for which the remnant remains bloated, $t_{bloat}$ is a few $\times10^{4}$ years, several hundred to a thousand times shorter than its likely survival time if it contracts onto the main sequence undisturbed, $t_{MS}$. Let the probability of the remnant being struck a second time while in its bloated state be $P_{bloat}\propto\sigma_{bloat}t_{bloat}$ and the probability of the remnant being struck again after contracting to the main sequence be $P_{MS}\propto\sigma_{MS}t_{MS}$. It follows from the rough cancellation of the factors by which $\sigma_{bloat}>\sigma_{MS}$ and by which $t_{bloat}<t_{MS}$ that, for any given remnant, $P_{bloat}\sim P_{MS}$. In stellar clusters where the probability of repeated collisions is low, this relation is unimportant since $P_{bloat}$ and $P_{MS}$ are both very small, but it becomes more interesting if stars are likely to suffer more than on collision.\\
\indent  We repeated the common--envelope analyses from Section 3.3 to
determine the likely outcome of an encounter between a bloated
collision remnant and a main--sequence star, again taking the
remnant's capture radius to be $1000$ R$_{\odot}$. For
$\alpha\lambda\gtrsim7$ or for any value of $\gamma$ within Nelemans
and Tout's preferred range, we found that such an encounter was
unlikely to lead to a merger, but instead would eject the remnant's
envelope, leaving a binary composed of the cores of the merger remnant
and of the second intruder star. In stellar clusters where
$t_{enc}\sim t_{bloat}$, $P_{bloat}$ becomes significant and it follows that roughly half of collision remnants will be struck a second time in their bloated state and disrupted, while roughly half will be struck again after contracting onto the main sequence. The results of Section 3.5 suggest that those remnants surviving to complete their main--sequence phase may explode as hypernovae.\\
\indent If we take the remnant's capture radius to be $1000$
R$_{\odot}$, its mass to be $10$ M$_{\odot}$ and set the encounter
timescale in Equation \ref{eqn:tenc} to the duration of the bloated
state, we find that the number density at which this occurs is a few$\times10^{7}$
pc$^{-3}$ (inserting instead the \textit{main--sequence} radius and
lifetime gives a similar result).\\
\indent In denser stellar systems ($n\sim10^{8}$
pc$^{-3}$) the population of remnants contracting onto the main sequence is depleted, since the majority of remnants suffer subsequent collisions while bloated and never reach the main sequence. Since the encounter timescale is a few times shorter than
the thermal timescale of the product of the first collision, merger
remnants will be struck several times while swollen. If the timescale on which the cores inspiral into the centre of the object and merger is comparable to or longer than the collision time, at any
one instant, the envelope of the remnant may contain three or four cores, forming a \textit{bag of cores}. This state of affairs
would lead to a complex few--body encounter, possibly expelling the
lower--mass core(s) while common--envelope evolution ejects the
envelope, leaving a binary composed of the two highest--mass cores, or alternatively leading to the merger of two of the cores
\citep{2004MNRAS.352....1F}. If, instead, the timescale for the cores to inspiral and merge is shorter than the collision timescale, each new impactor will probably merge with the object undisturbed by interactions with other impactors. The evolution of such systems will be the topic of a subsequent paper.\\
\indent \cite{2002ApJ...576..899P} and more recently
\cite{freitag_rasio} and \cite{freitag_gurkan} have suggested that
clusters of still higher number densities ($10^{9}-10^{11}$ pc$^{-3}$,
reached during core collapse) may host a runaway merger process in which tens of merger events lead to the formation of an object of mass $\sim10^{3}$ M$_{\odot}$ which subsequently explodes and leaves behind an intermediate--mass black hole. We do not think this scenario is very likely in practice. At these stellar densities, the encounter timescales are much shorter than the thermal timescale of the remnant of the initial merger, so collision products are pummelled by tens of impacts before they get the opportunity to swell on a thermal timescale. However, they will still swell on a dynamical timsecale, so later mergers may eject significant quantities of mass. In addition, massive stars have very powerful winds which drive mass--loss rates that can reach $10^{-4}$ M$_{\odot}$ yr$^{-1}$ even for `ordinary' OB or Wolf--Rayet stars \citep{1988A&AS...72..259D}. The wind from a star with a mass of $\sim1000$ M$_{\odot}$ is likely to be stronger still. It is therefore likely that the runaway buildup of mass becomes self--limiting, so that the rate at which the merger product gains mass by mergers is comparable to the rate at which mass is being ejected. Even if a very massive object does form, once collisions cease and the remnant approaches thermal equilibrium, it will swell to a vast radius and develop a very pronounced core/envelope structure, so that most of its mass is concentrated in a very small volume at its centre, the rest being distributed in a tenuous envelope. This structure, together with the tremendous luminosity that the core of such an object must have, will almost certainly lead to instabilities driving mass loss rates of at least  $\sim10^{-3}$ M$_{\odot}$ yr$^{-1}$ \citep{1993MNRAS.263..375G}.\\
\indent We summarise the probable fates of collision remnants in
stellar environments of increasing number density in Table
\ref{tbl:numdens}. In Figure \ref{fig:tree}, we explain schematically
the sequence of stellar encounters corresponding to each row in Table
\ref{tbl:numdens}. The stellar number density in clusters,
particularly in the cores of dense clusters, can increase dramatically
on timescales comparable to or shorter than massive stars'
main--sequence lifetimes due to relaxation--driven core--collapse and this evolution can be influenced by stellar collisions, since collisions can accelerate core collapse \citep{freitag_gurkan}. Conversely, the expulsion of gas be feedback from young stars can decrease the core stellar number density (e.g. \cite{2001MNRAS.321..699K})In addition, the number density in a given
cluster clearly varies spatially, generally being higher in the core
than at, for example, the half--mass radius. Table
\ref{tbl:numdens} and Figure \ref{fig:tree} are intended to show the
expected fate of stars in environments with the number densities
indicated. A given cluster may pass through some or all of these
density regimes as it evolves and may exhibit some or all of these
regimes at any given moment in time.\\
\begin{table*}
\begin{tabular}{|l|l|l|}
\hline
Case & Number density & Number of collisions per object and fate of collision products\\
 & (pc$^{-3}$) &\\
\hline
(0) & $n<10^{4}$ & Collisions rare - stars evolve normally.\\
(1) & $10^{4}<n<10^{6}$ & $0.1-10\%$ of stars will suffer single collisions. Products go on to explode as rapidly--rotating \\
 & & supernovae (RSN) or hypernovae if sufficiently massive ($\gtrsim20$ M$_{\odot}$).\\
(2) & $n\sim10^{7}$ & Most stars suffer $1-2$ collisions. About half of products likely to be struck again while \\
 & & swollen and to lose their envelopes, leaving a hard binary. The other half suffer a second\\
  & & collision on the main sequence and go on to explode as rapidly--rotating supernovae (RSN) or hypernovae\\
  & & if sufficiently massive ($\gtrsim20$ M$_{\odot}$).\\
(3) & $n\sim10^{8}$ & All stars likely to suffer several collisions. Collision remnants likely to be struck two or more \\
 & & times while in their swollen state. Interaction of the cores with each other and with the envelope \\
 & & likely to eject the lowest--mass cores as well as the envelope, leaving highest--mass cores in a hard \\
 & & binary.\\
(4) & $10^{9}<n<10^{11}$ & All stars likely to be involved in collisions. Collision timescale much shorter than remnants' thermal \\
 & & timescales so remnants have no opportunity to swell. Multiple mergers may build up an object \\
 & & of mass $\sim100$ $M_{\odot}$ after which mass buildup probably self--limited by winds.\\
\hline
\end{tabular}
\caption{Likely numbers of collisions and fates of collision products in stellar environments of increasing number density. Note that any given star cluster may pass through some or all of these regimes as it evolves and may exhibit several of these regimes at any given moment in time.}
\label{tbl:numdens}
\end{table*}

\section{Conclusions}
We have studied encounters between \ninesg and \onesg and between two \ninesg in stellar systems with velocity dispersions of $\sim10$ km s$^{-1}$. The capture radius for encounters between \ninesg and \onesg with this velocity dispersion is $\approx11$ R$_{\odot}$ and that for encounters between two \ninesg is $\approx18$ R$_{\odot}$.\\
\indent Using high-- and very high--resolution SPH simulations, we showed that encounters with small periastrons ($r_{p}\leqslant3$ R$_{\odot}$ in encounters between \ninesg and \ones, and $r_{p}\leqslant4$ R$_{\odot}$ in encounters between two \nines) resulted in mergers taking place on timescales of tens to hundreds of hours. By comparing results obtained at two very different numerical resolutions, we ascertained that such short merger times are not an artefact of numerical resolution.\\
\indent Encounters with larger periastrons led to captured systems whose orbits would probably shrink  over timescales of hundreds of hours to hundreds of years, leading to a phase of common envelope evolution. We studied the evolution of the common envelope systems resulting from the grind--down of the captured systems using the standard energy--conserving common--envelope formalism and the angular--momentum--conserving method of \cite{2004RMxAC..20...39N}. The energy--conserving formalism suggested that common--envelope encounters resulting from the grind--down of a captured system would result in a merger unless extreme values of the product $\alpha\lambda$ were invoked -- $\gtrsim20$ in encounters between \ninesg and \ones, and $\gtrsim100$ in encounters between two \nines. The results for the same encounters derived using the angular--momentum--conserving formalism were different and very sensitive to the chosen value of the $\gamma$ parameter. We again considered common--envelope evolution following the grind--down of a captured system. In encounters between \ninesg and \ones, we found that values of $\gamma\gtrsim1.65$ resulted in envelope--ejection without core merger. In encounters between two \nines, any value within Nelemans and Tout's preferred range resulted in envelope--ejection without core merger. We also found that more massive stars ($\approx50$ M$_{\odot}$), with their relatively more massive cores and less massive envelopes, are more vulnerable to envelope expulsion.\\
\indent We integrated our SPH simulations of the encounters between \ninesg and \onesg with periastrons $\leqslant3$ R$_{\odot}$ and those between two \ninesg with periastrons $\leqslant4$ R$_{\odot}$ until the stellar cores merged. These mergers ejected small quantities of mass, and, in mergers of \ninesg and \ones, the smaller star spiralled largely intact into the core of the larger star. The $9$ M$_{\odot}$ models used in our calculations had sufficient core hydrogen for $\approx2\times10^{6}$ yr of main sequence evolution. Merging these objects with \onesg delivered enough fresh hydrogen to the cores of the merger remnants to allow them to burn for another $\approx6\times10^{6}$ yr. We conclude that merger with a lower--mass object may extend the duration of a massive star's main--sequence phase by up to $\sim10\%$, although such an extension is likely to be smaller in reality, since we considered massive stars close to terminal age.\\
\indent The products of our collisions were rapidly rotating. The merger of two \ninesg with a periastron of $2$ R$_{\odot}$ produced an object with a core rotating sufficiently rapidly to make it a candidate hypernova/gamma ray burst progenitor according to the criterion given in \cite{2004ApJ...607L..17P}. Common--envelope evolution may expel some or all of the remnants' envelopes, so that their subsequent supernova/hypernova explosions are of Type Ib/c. Taking the typical stellar clusters in which massive stars are found to have stellar number densities of $\sim10^{4}$ pc$^{-3}$, we found that $\sim0.1\%$ of massive stars will suffer collisions, although only remnants with masses $\gtrsim20$ M$_{\odot}$ are massive enough to eventually explode as hypernovae. \cite{2004ApJ...607L..17P} estimate that $\sim0.1\%$ of massive stars explode as hypernovae. We showed that merger remnants possess sufficient core specific angular momentum after their formation to make them possible HNe/GRB progenitors, that the typical encounter timescales in massive stellar systems imply a collision rate similar to the hypernova rate and suggest that common--envelope evolution and winds may expel the envelopes of collision remnants. We conclude that the products of massive stellar collisions may contribute significantly to the observed hypernova rate.\\
\indent We studied further interactions of our collision products with main--sequence stars. The collision remnants swell on a thermal timescale, reaching radii large enough that an encounter with another star would result in a phase of common envelope evolution which expels the remnant's envelope. However, once the excess energy has been radiated away, an undisturbed remnant contracts (also on a thermal timescale) and becomes a main--sequence star. In stellar clusters with number densities of $\sim10^{7}$ pc$^{-3}$, about half of the collision products suffer a second collision back on the main sequence while the other half are struck whilst swollen and lose their envelopes, leaving a hard binary. In systems with number densities of $\sim10^{8}$ pc$^{-3}$, the collision product will suffer multiple further collisions whilst swollen, leading to a system in which several (i.e.$>$two) cores orbit within a common envelope. Interactions between the cores may eject all but the two most massive and drive off the envelope, leaving a hard binary, or alternatively lead to a merger of two of the cores. We will investigate this problem in detail in a subsequent paper. In denser systems ($n\gtrsim10^{9}$ pc$^{-3}$) the remnant of the first collision will be struck many times. \cite{2002ApJ...576..899P} and \cite{freitag_gurkan} and \cite{freitag_rasio} have suggested that such a runaway merger process will result in the formation of a very massive ($\sim1000$ M$_{\odot}$) object which eventually explodes and forms an intermediate--mass black hole. We do not think this scenario is likely as hydrodynamic swelling of the remnant, making it vulnerable to mass--ejection in subsequent collisions, and strong stellar winds would probably make the mass buildup self limiting when the merger product acquires a mass of a few hundred solar masses. In addition, the collision remnant may become `transparent' to subsequent imapctors \citep{1967ApJ...150..163C}.Even if a $\sim1000$ M$_{\odot}$ object does form, we suggest that winds and instabilities are likely to result in mass losses $>10^{3}$ M$_{\odot}$ yr$^{-1}$. By the time the object reaches the end of its main sequence phase, it is likely to have lost most of its mass and its supernova explosion will not produce an unusually massive black hole.\\

\section{Acknowledgements}
JED acknowledges support from the Leicester PPARC Rolling Grant. MBD acknowledges the support of a research fellowship awarded by the Royal Swedish Academy of Sciences. Some of the calculations presented in Section 3 were performed using the UK Astrophysical Fluids Facility (UKAFF). We thank an anonymous referee for very detailed and useful comments.

\bibliography{myrefs}

\label{lastpage}

\end{document}